\def\b{\begin{equation}}
\def\e{\end{equation}}
\def\l{\left}
\def\r{\right}
\documentclass[prl,aps,twocolumn, showpacs, showkeys, floats]{revtex4}

\begin{document}

\title
{Macroscopic form of the first law of thermodynamics for an adiabatically evolving {\em non-singular} self-gravitating fluid}

\author{Abhas Mitra}

\email {amitra@barc.gov.in}
\affiliation { Theoretical Astrophysics Section, Bhabha Atomic Research Centre, Mumbai, India}


\date{\today}






\date{\today}
\begin{abstract}
We emphasize that the pressure related  work appearing in a general relativistic first law of thermodynamics should involve {\em proper volume element} rather than coordinate volume element. This point is highlighted by considering both local energy momentum conservation equation as well as particle number conservation equation. It is also emphasized that we are considering here a {\em non-singular} fluid governed by purely classical general relativity. Therefore, we are not considering here any semi-classical or quantum gravity which apparently suggests thermodynamical properties even for a (singular) black hole. Having made such a clarification, we formulate a global first law of thermodynamics for an adiabatically evolving spherical perfect fluid. It may be verified that such a global first law of thermodynamics, {\em for a non-singlar fluid}, has not been formulated earlier. 
\end{abstract}

\maketitle

\section {Introduction}
We shall consider here the adiabatic evolution of a self-gravitating {\em non-singular} perfect fluid from the view point of first law of thermodynamics. By non-singular, we mean that we would not consider any black hole or naked singularity here. Further we shall analyze the problem in the framework of {\em classical} general relativity (GR) rather than any semi-classical or quantum gravity theory. In fact, it may be borne in mind that from the view point of {\em classical} GR, a Schwarzschild black hole is characterized by a single number which is its gravitational mass $M_0$. Accordingly its entropy should be $S_{BH}(classical)= k\ln 1 =0$. Also its temperature should be $T=0$ because it cannot radiate.

 It hardly needs to be reiterated that the first law thermodynamic (FLT) is a statement of total energy conservation. In its usual macroscopic form it says
that the change in ``internal energy'' of a system having a fixed number of particles $N$ is given by
\b
dU = dQ - pdV
\e
where $dQ$ is the heat/radiation injected {\em into} the system and $p dV$ is amount of work done {\em by} the system on its ``surroundings''. And if it would be assumed that  there is  neither any external heat injection nor any internal heat generation, then one would have
$d Q =0$. Thus, for a supposed adiabatic evolution of a perfect fluid, $dQ=0$.
 
 Now from purely Newtonian thermodynamics, if one would shift to  special  theory of relativity (STR), the internal energy should be replaced by the total mass energy $E= M_i c^2$, where $M_i$ is the inertial mass of the system in its rest frame. Here $c$ is the speed of light. However, we will mostly take $G=c=1$. Since the spacetime is flat in STR, the volume element involved in Eq.(1), $dV$,  continues to imply just the coordinate volume element: $dV = dx dy dz= R^2 dR \sin \theta d\theta d\phi$, where $\theta$ and $\phi$ are polar angles.  
In this case,
FLT becomes
\b
dE + p dV=0
\e
However if one would consider general theory of relativity (GTR), spacetime is curved and one must consider proper volume $dv = \sqrt{h} ~dV$ where $h$ is the negative of the spatial part of the metric determinant $g$\cite{1}, i.e.,
 \b
 -g = g_{00} (-h)
 \e
 It must be so because in GR, (1) one can choose arbitrary coordinate system so that many forms of $dV$ could be {\em physically meaningless as a physical volume element} and also because (2) {\em in the presence of gravity the 3-space gets curved}. Further, in curved spacetime, energy measured locally would be different from the one measured by a distant observer. And one needs to account for such a gravitational redshift factor too for a consistent macroscopic formulation of FLT.

   Having made this preliminary background,
 let us recall that   energy momentum is conserved at a given spacetime point in GR.   Accordingly, in general relativity, one should first obtain a local
 microscopic FLT as has been extensively done by Tolman\cite{1}:
 \b
 d(\delta E) + p d(\delta v) =0
 \e
 Here one considers an infinetisimal proper volume element $\delta v$ and applies FLT. It is likely that, in the limit $\delta v \to 0$, the effect
 of global self-gravitational energy vanishes and one can indeed write
 \b
 \delta E = \rho ~\delta v
 \e
 Later, we shall rigourously verify that GR indeed yields such a microscopic version of FLT.
 
 Finally we will arrive at the macroscopic local form of FLT (MFLT) for an adiabatically evolving spherical fluid by considering the effect of spacetime curvature.

 \section{Purely Local FLT}
 
 As is well known, the local energy momentum conservation is represented by the vanishing of the covariant divergence of the energy momentum tensor\cite{1,2,3,4}
 $T^{ij}$:
 \b
 {T^{ij}}_{;j} =0
 \e
 In terms of density of energy momentum tensor
 \b
 {\cal T}^{ij} = \sqrt{-g}~ T^{ij}
 \e
 the energy momentum consevation can be expressed as\cite{1}
 \b
 {\partial {\cal T}^{ij}\over \partial x^i} - {1\over 2} {\cal T}^{kl} {\partial g_{kl} \over \partial x^j} =0
 \e
 Suppose we are considering a spacetime metric of signature $-2$ and a perfect fluid with energy momentum tensor:
 \b
T^{ij} = (\rho +p) u^i u^j - p g^{ij}
\e
where $u^i$ the fluid 4-velocity, $\rho$ is density and $p$ is the isotropic pressure. If we would use comoving coordinates, $u^i$ has only the non-vanishing temporal component:
 \b
 u^i = e^{-\nu/2} \delta^i_0
 \e
 where $g_{00} =e^\nu$. Then one finds
 \b
 T_{00}  = \rho
 \e
 while
 \b
 T_{11} = T_{22} = T_{33} = -p,
 \e
Further,
 \b
 T^{00} = \rho g^{00},
 \e
 and
 \b
 T^{\mu \nu} = -p g^{\mu \nu}
 \e
 Here Greek indices run from $1,2,3$.
 Note, had there been any heat flow, $T^{ij}$ would have been non-diagonal. While the spatial part of Eq.(8)  gives ``momentum'' conservation, the temporal
 part yields ``energy'' conservation. Accordingly, we consider $j =0$ in Eq.(8) to obtain
 \b
 {\partial {\cal T}^{i0}\over \partial x^i} - {1\over 2} {\cal T}^{kl} {\partial g_{kl} \over \partial t} =0
 \e
 Using Eqs.(13) and (14), we simply the foregoing Eq. as
 \b
 {\partial\over \partial t} (\sqrt{-g} \rho)  + {1\over 2} p \sqrt{-g} {1\over g_{\mu \nu}} {\partial g_{\mu \nu} \over \partial t}
 - {1\over 2} \rho \sqrt{-g} {1\over g_{00}} {\partial g_{00} \over \partial t} =0
 \e
 Now using the fact $-g = (- h) g_{00}$, 
 \b
 {1\over x} {\partial x \over \partial t} = {\partial \ln x\over \partial t}
 \e
 and
 \b
 {\partial x\over \partial t} = 2\sqrt{x} {\partial \sqrt{x} \over \partial t}
 \e
Eq.(16) gets reduced to
 \b
 {\partial\over \partial t} (\sqrt{h} \rho) + p {\partial \sqrt{h} \over \partial t} =0
 \e
 Since the  comoving proper volume element is $\delta v = \sqrt{h} \delta^3 x$ and Eq.(19) concerns partial derivative with respect to $t$, we
 can multiply it by $\delta^3 x$ to obtain
 \b
 {\partial\over \partial t} ( \rho \delta v) + p {\partial \delta v \over \partial t}=0
 \e
 And this can immediately be interpreted in terms of Eqs.(4) and (5) as equation for local energy conservation. Note, unlike Tolman\cite{1} (p. 379), we derived  Eqs.(19) and (20) without restricting $g_{00} =1$.
 However, eventually, no $g_{00}$ appears here because both $\rho \delta v$ and $p \delta v$ represent
 {\em locally} measured energies. This clearly shows that, in curved spacetime, the pressure related work $dw$ must be formulated in terms of the proper
 volume element $dv$ rather than the flat volume element $dV$. It may be also noted that both $\rho \delta v$ and $p \delta v$ are energies measured by the {\em same local comoving frame}. Therefore Eq.(4) indeed is a correct form of local GR FLT for adiabatic evolution. No self-gravitational energy appears here because of Principle of Equivalence (POE) by which gravitational energy may be made to vanish {\em (only) at a given point}.

Now, if we define
\b
 {\cal \theta} = { \partial  (\delta v) /\partial t \over (\delta v)} 
 \e
as the volume expansion scalar, Eq.(20) may be also expressed as
\b
{\dot \rho} + (\rho +p) {\cal {\theta}}  = 0
\e
where an overdot represents differentiation w.r.t. comoving time $t$  Note, however, that the more formal definition of the expansion scalar however is\cite{2,5}
 \b
 \theta = {u^i}_{;i}
 \e
 \subsection {Particle Number Conservation}
 In the following, we shall explore the close link between microscopic FLT and conservation of total number of baryons or particles. 
 If a comoving fluid element contains a fixed number of partcles
 $\delta N$ and has a proper volume $\delta v$ then\cite{2} 
 \b
 \delta N = n \delta v
 \e
 is conserved and accordingly
 the comoving temporal derivative of $\delta N$ must be zero:
 \b
 {\partial {\delta N}\over \partial t} =0
 \e
 while a more formal statement for particle number conservation\cite{2,3,4} is:
 \b
 {(n u^i)}_{;i} =0
 \e
 Note, Eq.(25) leads to
 \b
 {\dot n \over n} + {\cal {\theta}} =0; \qquad {\cal {\theta}} = - {\dot {n}\over n}
 \e
 
 Though  $\sqrt{g_{00}}$  does not appear  here,  one can nonethess divide the above equation by $\sqrt{g_{00}}$ and interpret the time differentiation as w.r.t. proper time $\tau$.

By using Eq.(27) in (22), we have
\b
{\dot \rho} - (\rho +p) {{\dot n} \over n} =0
\e
And this may be rewritten as
 
 \b
 d\l({\rho \over n}\r) + p d\l({1\over n}\r) = 0
 \e
Note that, although the foregoing Eq. is supposed to be a GTR equations, no spacetime curvature term explicitly appears there. Consequently, one obtains
 exactly the same equations by applying STR as well\cite{4}. The reason for this could be that, in the presence of gravity, all evoltion must be dissipative and non-adiabatic\cite{6}. Conversely, {\em all strictly adiabatic evolution may fall in the STR domain}.

If now, we consider, $N =1$, i.e., the volume $\delta v$ contains a single particle and
 \b
 \delta v = \l({1\over n}\r)
 \e
 we will obtain the  more familiar form of (local) FLT:
 \b
 d (\rho \delta v) + p d(\delta v) =0
 \e
Thus, for adibatic motion, the FLT is closely linked with particle number conservation, and in GR, both of which must be formulated in terms of {\em proper volume element rather than coordinate volume element}.

 \section{Adiabatic Evolution Equation of a Spherically Symmetric  Fluid}
 
 Now we consider the specific case of a spherically symmetric adiabatic evolution by adopting the comoving spherically symmetric metric:
 \b
 ds^2 = e^\nu dt^2 - e^\lambda dr^2 - R^2 (d\theta^2 + \sin^2 \theta d\phi)
 \e
 Here $R = R(r,t)$ is the areal coordinate, $r$ is comoving radial coordinate. Physically $r$ is proportional to the number of particles
 enclosed by a given mass shell; $\lambda = \lambda(r,t)$; $\nu =\nu(r,t)$. It is also conventional to write
 \b
 R^2 = e^\mu
 \e
 so that
 \b
 {\dot \mu} = {2 {\dot R} \over R}
 \e 
For the metric (32), one has
\b
\sqrt{h} = R^2 e^{\lambda/2} \sin \theta
\e
and, the proper volume element is
\b
 dv = \sqrt{h}  dR ~d\theta ~d\phi
\e
As we use Eqs.(34) and (35) in Eq.(19), we obtain\cite{3}

 \b
 {\dot \rho} + ( {\dot \mu} + {1\over 2} {\dot \lambda}) (p + \rho)  =0
 \e
Thus, for the spherical symmetric case, by comparing Eqs.(22) and (37), we find that
\b
{\cal \theta} = {\dot \mu} + {1\over 2} {\dot \lambda}
\e
Let us also recall that the gravitational mass or Misner-Sharp mass of a section of the fluid in this case is given by\cite{5}
\b
M(r,t) = \int_0^r 4 \pi R^2 \rho ~R' dr
\e

 \section{Macroscopic  Ist Law of Thermodynamics}
  Now we shall consider application of FLT for an adibatically evolving spherical perfect fluid. This will be different from the local FLT which is strictly valid only at a point. If one would consider the motion of  a shell with {\em fixed number} of particles, one would have
 \b
 dR= dR|_{r=r} = {\dot R} ~dt
 \e
 And the coordinate volume swept by this shell is
 \b
 dV = 4 \pi R^2 dR|_{r=r} = 4\pi R^2 ~{\dot R} ~ dt
 \e
 
 Accordingly, the flat spacetime FLT should be obtainable from
 \b
 d E +   p~dV = dE +  4 \pi p R^2  ~ {\dot R} dt  =0
 \e
 
 But in curved spacetime  one must  consider the
 swept up {\em proper} volume element \cite{1} rather than {\em flat} volume element:
 \b
 dv = \sqrt{-g_{rr}} ~dV =  4 \pi  R^2  e^{\lambda/2}~ {\dot R} ~dt
 \e
 And if the spacetime has no gravitational redshift,  the macroscopic form of  adiabatic FLT would
 simply be
 \b
 d E +  ~ p~ dv =0
 \e
 Using Eq(43)., this becomes
 \b
 d E +  4 \pi p  R^2   e^{\lambda/2}~ {\dot R} ~dt =0
 \e

 But if the total energy $E$ is measured by a distant intertial observer $S_\infty$, note that the local $p dv$ work will appear to $S_\infty$ as\cite{6}
 \b
 dW_\infty = \sqrt{g_{00}}~ dW =e^{\nu/2} ~p~dv
 \e
 owing to the gravitational redshift.
 Thus, for a purely adiabatic case, the macroscopic GR FLT would assume the form
 \b
  dE +   4 \pi p R^2 ~  e^{(\lambda +\nu)/2}~ {\dot R} dt  =0
 \e
But by POE, we must have, $M_i = M$ and $E=M$ so that the above equation becomes
\b
  dM +   4 \pi p R^2 ~  e^{(\lambda +\nu)/2}~ {\dot R} dt  =0
 \e
 Note that, indeed $M$ is the ADM mass and is measured by a distant observer in an assumed asymptoptically flat spacetime.

 \section{Summary}
We showed that

$\bullet$ In GR, both local energy conservation and particle number conservation involve proper volume element rather than coordinate/flat volume element. It must be so because {\em in curved spacetime, coordinate volume element could be meaningless}.

$\bullet$ Accordingly, we formulated here the First Law of Thermodynamics for an adiabatically evolving fluid which obeys particle number conservation

 It is again reminded that, as per classical GR, a Schwarzschild black hole has zero entropy, and the ad-hoc idea of  BH thermodynamics can be recoinciled only if BHs have zero surface\cite{7} area or by yet unknown ad-hoc semi-classical gravity. In contrast, we are concerned here with mundane but solid classical GR. It will be shown in future work that the {\em classical GR}  FLT obtained here has important consequences for (classical) gravitational collapse and cosmology.
 
 \section{Acknowledgements}
 The author thanks the anonymous 2nd and 3rd referee for their honest reviews and suggestions for this manuscript.
 


\end{document}